\documentclass[5p,twocolumn,sort&compress,times,reversenotenum]{elsarticle}
\usepackage[colorlinks=true, pdfstartview=FitV,
bookmarks=true, bookmarksnumbered=true, breaklinks]{hyperref}
\usepackage{color,graphicx}
\definecolor{deeppink1}{rgb}{1.0000, 0.0784, 0.5765}
\definecolor{darkviolet}{rgb}{0.5804, 0.0000, 0.8275}
\definecolor{yellowgreen}{rgb}{0.6039, 0.8039, 0.1961}
\hypersetup{linkcolor=deeppink1, citecolor=darkviolet, urlcolor=yellowgreen}

\usepackage{amsmath,amssymb,bm,color,longtable,mathrsfs,slashed,comment}
\usepackage{fancyhdr}
\begin{document}

\begin{frontmatter}
\title{Casimir effect for lattice fermions}
\author[1,2]{Tsutomu~Ishikawa}
\ead{tsuto@post.kek.jp}
\author[3]{Katsumasa~Nakayama}
\ead{katsumasa.nakayama@desy.de}
\author[4]{Kei Suzuki}
\ead{k.suzuki.2010@th.phys.titech.ac.jp}

\address[1]{Graduate University for Advanced Studies (SOKENDAI), Tsukuba, 305-0801, Japan}
\address[2]{KEK Theory Center, Institute of Particle and Nuclear Studies, High Energy Accelerator Research Organization (KEK), Tsukuba, 305-0801, Japan}
\address[3]{NIC, DESY Zeuthen, Platanenallee 6, 15738 Zeuthen, Germany}
\address[4]{Advanced Science Research Center, Japan Atomic Energy Agency (JAEA), Tokai, 319-1195, Japan}

\begin{abstract}
We propose a definition of the Casimir energy for free lattice fermions.
From this definition, we study the Casimir effects for the massless or massive naive fermion, Wilson fermion, and (M\"obius) domain-wall fermion in $1+1$ dimensional spacetime with the spatial periodic or antiperiodic boundary condition.
For the naive fermion, we find an oscillatory behavior of the Casimir energy, which is caused by the difference between odd and even lattice sizes.
For the Wilson fermion, in the small lattice size of $N \geq 3$, the Casimir energy agrees very well with that of the continuum theory, which suggests that we can control the discretization artifacts for the Casimir effect measured in lattice simulations.
We also investigate the dependence on the parameters tunable in M\"obius domain-wall fermions.
Our findings will be observed both in condensed matter systems and in lattice simulations with a small size.
\end{abstract}
%\begin{keyword}
%\end{keyword}
\end{frontmatter}

\thispagestyle{fancy}
\rhead{DESY 20-094, KEK-TH-2220}
%\cfoot{}
\renewcommand{\headrulewidth}{0.0pt}
%\pagestyle{empty}

%%%%%%%%%%%%%%%%%%%%%%%%%%%%%%%%
\section{Introduction} \label{Sec_1}
%%%%%%%%%%%%%%%%%%%%%%%%%%%%%%%%
The Casimir effect~\cite{Casimir:1948dh} is known as negative energy and attractive force caused by a zero-point energy shift of photon fields between two parallel plates.
It was first predicted in 1948~\cite{Casimir:1948dh} and was experimentally discovered after fifty years~\cite{Lamoreaux:1996wh}.
The original Casimir effect is physics related to photon fields, which is perturbatively described in quantum electrodynamics (QED), but a similar concept can be applied to any field such as scalar fields, fermion fields, and other gauge fields.

Nowadays, lattice simulations are utilized as a tool for controllably studying various quantum phenomena, and Casimir(-like) effects {\it on the lattice} for not only the $U(1)$ gauge theory~\cite{Pavlovsky:2009kg,Pavlovsky:2010zza,Pavlovsky:2011qt} but also scalar field theories~\cite{Chernodub:2019kon}, and nonperturbative theories such as the compact $U(1)$ gauge theory~\cite{Pavlovsky:2009mt,Chernodub:2016owp,Chernodub:2017mhi,Chernodub:2017gwe} and Yang-Mills theories with $SU(2)$~\cite{Chernodub:2018pmt,Chernodub:2018aix} and $SU(3)$~\cite{Kitazawa:2019otp} gauge fields were measured.
Furthermore, in the future, the Casimir effect in lattice gauge theories coupled to dynamical fermions, such as QED and quantum chromodynamics (QCD) will draw attention.
In particular, the {\it QCD Casimir effect} contains complicated contributions from not only perturbative quarks and gluons but also nonperturbative phenomena such as the spontaneous chiral-symmetry breaking, confinement, instantons, and effects from hadron degrees of freedoms (for related works from fermionic effective models, see Refs.~\cite{Kim:1987db,Song:1990dm,Song:1993da,Kim:1994es,Vshivtsev:1995xx,Vdovichenko:1998ev,Vshivtsev:1998fg,Braun:2004yk,Braun:2005gy,Braun:2005fj,Abreu:2006pt,Ebert:2008us,Palhares:2009tf,Abreu:2009zz,Abreu:2010zzb,Hayashi:2010ru,Ebert:2010eq,Braun:2010vd,Abreu:2011zzc,Ebert:2011tt,Braun:2011iz,Flachi:2012pf,Flachi:2013bc,Tiburzi:2013vza,Tripolt:2013zfa,Phat:2014asa,Ebert:2015vua,Almasi:2016zqf,Flachi:2017cdo,Nitta:2017uog,Abreu:2017lxf,Ishikawa:2018yey,Inagaki:2019kbc,Xu:2019gia,Abreu:2019czp,Ishikawa:2019dcn,Abreu:2019tnf,Abreu:2020uxc}).\footnote{The first example of the Casimir effects for massless quarks and gluons was introduced in the context of the MIT bag model by Johnson~\cite{Johnson:1975zp}.}
However, a discretized lattice-field theory is different from the original continuum theory, so that, to precisely interpret physics measured on the lattice, we also need the understandings of the Casimir effects for lattice field theories. 
 
The purpose of this Letter is to define the Casimir energy for free lattice fermions for the first time.\footnote{There are a few analytical works about the Casimir effects for lattice scalar fields~\cite{Actor:1999nb,Pawellek:2013sda}.}
The Casimir energy for continuous degrees of freedom can be derived by dealing with divergent zero-point energy.
On the other hand, the zero-point energy for lattice degrees of freedom is not divergent because of the lattice regularization.
In this sense, the definition of the Casimir energy on the lattice is not trivial.
Furthermore, our motivations for this study are as follows:
\begin{enumerate}
\item It will be important for deeply understanding the elemental properties of lattice fermion actions.
This is because the Casimir effect might be sensitive to the properties of lattice fields in the ultraviolet (UV) region.
In the UV region, the properties of lattice fermions are deformed by a nonzero lattice spacing $a$, where $1/a$ gives a UV cutoff scale in momentum space, so that the physics would different from that in the continuum theory.
In this sense, the comprehensive examination of the phase structure of a fermion action in finite (especially, small) volume will be important, which is similar to theoretical studies at finite temperature and/or density.
\item It will be useful for the estimate or interpretation of discretization artifacts contaminating in lattice simulations.
In this work, for simplicity, we focus on only the free (noninteracting) fermions, but the studies of qualitative properties would be useful also for lattice simulations with interacting fermions such as lattice QCD.
\item It can be related to the Casimir effects in similar systems in condensed matter physics.
For example, Wilson fermion-like dispersion relations appear in Dirac semimetals~\cite{Zhang:2009zzf,Wang:2012,Wang:2013}.
Also, the domain-wall fermions are well known as an analogy to zero-mode Dirac fermions realized on the surface of topological insulators \cite{Qi:2005,Qi:2008}.
In this sense, this study is not limited to theoretical interests, and it can also provide us motivations for future tabletop experiments in condensed matter.
In such situations, we can experimentally observe Casimir effects for (Dirac-like) lattice fermions.
\end{enumerate}

When we naively formalize fermion fields on the lattice~\cite{Wilson:1974sk}, we confront the so-called doubler problem, which is known as the Nielsen-Ninomiya no-go theorem~\cite{Nielsen:1980rz,Nielsen:1981xu}.
To evade from the doubler problem, one has to introduce ``improved" fermion actions such as the Wilson fermions~\cite{Wilson:1975,Wilson:1977} and domain-wall (DW) fermions~\cite{Kaplan:1992bt,Shamir:1993zy,Furman:1994ky}.
Although the Wilson fermion breaks the chiral symmetry, it has been broadly used in the various simulations of QCD.
On the other hand, the DW fermion formulation realizes the chiral symmetry on the lattice and has been successfully applied to the investigation of various physics. 
The M\"obius domain-wall (MDW) fermion~\cite{Brower:2004xi,Brower:2005qw,Brower:2012vk} is an improvement of the DW fermion using a real M\"obius transformation of the Dirac operator.

The contents of this Letter are organized as follows.
In Sec.~\ref{Sec_2}, we give a definition of Casimir energy for lattice fermions.
As an example of applications of this definition, in Sec.~\ref{Sec_3}, we investigate the properties of the Casimir energy for the naive lattice fermion.
Here, we will find a characteristic oscillation of Casimir energy.
In Secs.~\ref{Sec_4} and~\ref{Sec_5}, we study the Casimir energies for the Wilson and overlap fermions, respectively.
Section~\ref{Sec_6} is devoted to our conclusion and outlook.

%%%%%%%%%%%%%%%%%%%%%%%%%%%%%%%%
\section{Definition of Casimir energy on the lattice} \label{Sec_2}
%%%%%%%%%%%%%%%%%%%%%%%%%%%%%%%%
In this section, we give a definition of the Casimir energy for lattice fermions.
In this Letter, as a situation with the Casimir effect, we consider only the compactification of one spatial dimension.
Then only the first component $p_1$ of the spatial momentum of a fermion is discretized: $p_1 \to p_1(n)$, where $n$ is the label of discretized levels.
The other spatial components (e.g., $p_2$ and $p_3$ in three spatial dimensions) are not discretized.
Moreover, we can choose two types of definitions for the temporal component ($p_4$ in Euclidean space or $p_0$ in Minkowski spacetime): (i) the temporal component is not latticized, and (ii) the temporal component is latticized.
The definition (i) corresponds to materials with a small size in condensed matter systems, where the energy $p_0$ of a fermion is not latticized.
On the other hand, the definition (ii) corresponds to numerical lattice simulations with the temporal direction, where the (Euclidean) temporal component is also artificially latticized.
The situations (i) and (ii) share some properties of the Casimir effect, but the detail is slightly different.
In this Letter, we apply only the definition (i).
For the case (ii), see our future studies~\cite{Ishikawa:2020}.

First, we define the energy-momentum dispersion relation of fermions on the lattice.
This is defined by the combination of the Dirac operators.
In the $3+1$ dimensions,
\begin{align}
aE (ap) = a \sqrt{ D_k^\dagger D_k }, \label{eq:def_disp}
\end{align}
with the lattice spacing $a$, where $k=1,2,3$ is the index of the spatial component.
Note that this representation is equivalent to that extracted from the pole structure of the propagator of lattice fermions.

When we set the lattice size (or dimensionless volume) for a compactified spatial direction as $N=L/a$, the spatial momentum for this direction with the periodic and antiperiodic boundary conditions is discretized as
\begin{align}
ap_1 \to ap_1^\mathrm{P} (n) = \frac{2 n \pi}{N}, \ \ \ ap_1 \to ap_1^\mathrm{AP} (n) = \frac{(2n+1) \pi}{N},
\end{align}
respectively.
The label $n$ is an integer ($n =0, \pm 1,2, \cdots$).
When we choose the Brillouin zone for the three spatial momenta as $-\pi < ap_k \leq \pi$ or $0 \leq ap_k < 2\pi$, and then the lower and upper bounds of $n$ are determined by the Brillouin zone and the boundary condition.
%\begin{align}
%-\pi < ap_k \leq \pi &\to -\frac{N}{2} <n^\mathrm{P} \leq \frac{N}{2}, \label{eq:BZ1_n_P}\\
%-\pi < ap_k \leq \pi &\to -\frac{N}{2}-\frac{1}{2} <n^\mathrm{AP} \leq \frac{N}{2}-\frac{1}{2}, \label{eq:BZ1_n_AP} \\
%0 \leq ap_k < 2\pi   &\to  0 \leq n^\mathrm{P} < N, \label{eq:BZ2_n_P} \\
%0 \leq ap_k < 2\pi   &\to -\frac{1}{2} \leq n^\mathrm{AP} < N-\frac{1}{2},  \label{eq:BZ2_n_AP}
%\end{align}
%where $N$ and $n$ should be an integer, so that the range of $n^\mathrm{AP}$ in Eq.~(\ref{eq:BZ2_n_AP}) is practically $0 \leq n^\mathrm{AP} < N$.
Note that the Casimir energy defined below does not depend on the choice of the Brillouin zone.

By discretization of the momentum $p_1$, the zero-point energy (per area) is redefined as~\footnote{
The original zero-point energy in the three dimensional space is defined as
\begin{align}
a\hat{E}_0(N\to\infty) = - \frac{V}{a^3} c_\mathrm{deg} \int \frac{d^3ap}{(2\pi)^3} aE (ap), \nonumber
\end{align}
where $V= L \times A$ with the two-dimensional area $A$.
In this sense, Eq.~(\ref{eq:zeropoint_ene}) is the zero-point energy divided by the area.
}
\begin{align}
aE_0(N\to\infty) &= -N c_\mathrm{deg} \int \frac{d^3ap}{(2\pi)^3} aE (ap) \nonumber\\
\to aE_0(N) &= - c_\mathrm{deg} \int \frac{d^2ap_\perp}{(2\pi)^2} \sum_n aE_n (ap_\perp,N), \label{eq:zeropoint_ene}
\end{align}
where $c_\mathrm{deg}$ is the degeneracy factor, such as spin of fermion, and we set $c_\mathrm{deg}=1$ throughout this Letter.
The minus sign comes from the property of fermions.
The factor of $\frac{1}{2}$ from the zero-point energy and the factor of $2$ from the particle and antiparticle cancel out each other.

Here, we define the Casimir energy for lattice fermions (with one compactified space in the $3+1$ dimensional spacetime) as the difference between $aE_0(N)$ and $aE_0(N\to\infty)$:
\begin{align}
& aE_\mathrm{Cas}^\mathrm{4d} \equiv aE_0(N) - aE_0(N\to\infty) \label{eq:def_cas_4d} \\
& = c_\mathrm{deg} \int \frac{d^2ap_\perp}{(2\pi)^2} \left[ -\sum_n aE_n(ap_\perp,N) + N \int_\mathrm{BZ} \frac{dap_1}{2\pi} aE(ap) \right],  \nonumber
\end{align}
where the integral with respect to $p_1$ runs over the defined Brillouin zone.
Application to other dimensions is straightforward.
For example, in the $1+1$ dimensional spacetime,
\begin{align}
aE_\mathrm{Cas}^\mathrm{2d} \equiv c_\mathrm{deg} \left[ -\sum_n aE_n(N) + N \int_\mathrm{BZ} \frac{dap_1}{2\pi} aE(ap) \right]. \label{eq:def_cas}
\end{align}

%%%%%%%%%%%%%%%%%%%%%%%%%%%%%%%%
\section{Casimir energy for naive fermion} \label{Sec_3}
%%%%%%%%%%%%%%%%%%%%%%%%%%%%%%%%
First, we study the Casimir effect for the naive lattice fermion.
The (dimensionless) Dirac operator of the naive lattice fermion with a mass $m_f$ in momentum space is defined as
\begin{align}
aD_\mathrm{nf} \equiv i \sum_k \gamma_k \sin ap_k + am_f,
\end{align}
where $\gamma_k$ is the gamma matrix.

From Eq.~(\ref{eq:def_disp}), we can evaluate the dispersion relation:
\begin{align}
a^2 E_\mathrm{nf}^2(ap) = \sum_k \sin^2 ap_k + (am_f)^2.
\end{align}
From Eq.~(\ref{eq:def_cas}), we can calculate the Casimir energy.
Here, the integration with respect to $ap_\perp$ is limited to the first Brillouin zone, so that the Casimir energies are determined without any divergence.
Then we can numerically evaluate the Casimir energy.
In order to get the analytic formulas of the Casimir energy, one also can utilize a mathematical technique such as the Abel-Plana formulas~\cite{Ishikawa:2020}.

For the naive fermion with $m_f=0$ in the $1+1$ dimensional spacetime, we can get the exact formulas:
\begin{align}
aE_\mathrm{Cas}^\mathrm{2d,nf,P} &= \frac{2N}{\pi} - \cot \frac{\pi}{2N}  & (N=\mathrm{odd}), \label{eq:nf_exact1} \\
aE_\mathrm{Cas}^\mathrm{2d,nf,P} &= \frac{2N}{\pi} -  2\cot \frac{\pi}{N} & (N=\mathrm{even}), \label{eq:nf_exact2} \\
aE_\mathrm{Cas}^\mathrm{2d,nf,AP} &= \frac{2N}{\pi} - \cot \frac{\pi}{2N} & (N=\mathrm{odd}), \label{eq:nf_exact3} \\
aE_\mathrm{Cas}^\mathrm{2d,nf,AP} &= \frac{2N}{\pi} - 2\csc \frac{\pi}{N} & (N=\mathrm{even}), \label{eq:nf_exact4}
\end{align}
where we distinguished the solutions into the odd lattice ($N=\mathrm{odd}$) and even lattice ($N=\mathrm{even}$). 

When we expand these formulas by $a/L$ (or equivalently $1/N$) to see the series of a small lattice spacing $a$, we obtain
\begin{align}
E_\mathrm{Cas}^\mathrm{2d,nf,P} &= \frac{\pi}{6L} +\frac{\pi^3 a^2}{360L^3} + \mathcal{O}(a^4)   &(N=\mathrm{odd}), \label{eq:nf_aexp1} \\
E_\mathrm{Cas}^\mathrm{2d,nf,P} &= \frac{2\pi}{3L} +\frac{2\pi^3 a^2}{45L^3} + \mathcal{O}(a^4)  &(N=\mathrm{even}),  \\
E_\mathrm{Cas}^\mathrm{2d,nf,AP} &= \frac{\pi}{6L} +\frac{\pi^3 a^2}{360L^3} + \mathcal{O}(a^4)  &(N=\mathrm{odd}),  \\
E_\mathrm{Cas}^\mathrm{2d,nf,AP} &= -\frac{\pi}{3L} -\frac{7\pi^3 a^2}{180L^3} + \mathcal{O}(a^4)&(N=\mathrm{even}). \label{eq:nf_aexp4}
\end{align}

\begin{figure}[t!]
    \begin{minipage}[t]{1.0\columnwidth}
        \begin{center}
            \includegraphics[clip, width=1.0\columnwidth]{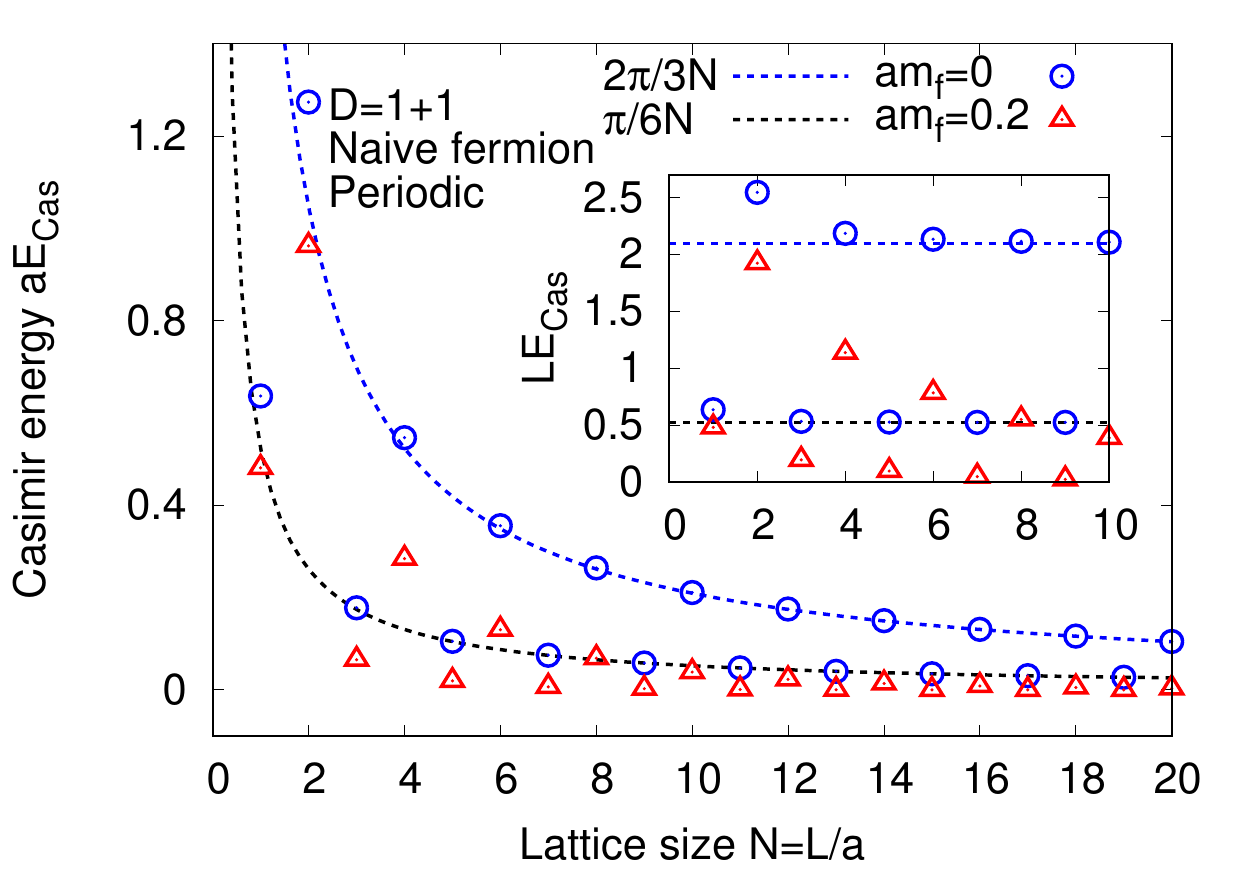}
        \end{center}
    \end{minipage}
    \begin{minipage}[t]{1.0\columnwidth}
        \begin{center}
            \includegraphics[clip, width=1.0\columnwidth]{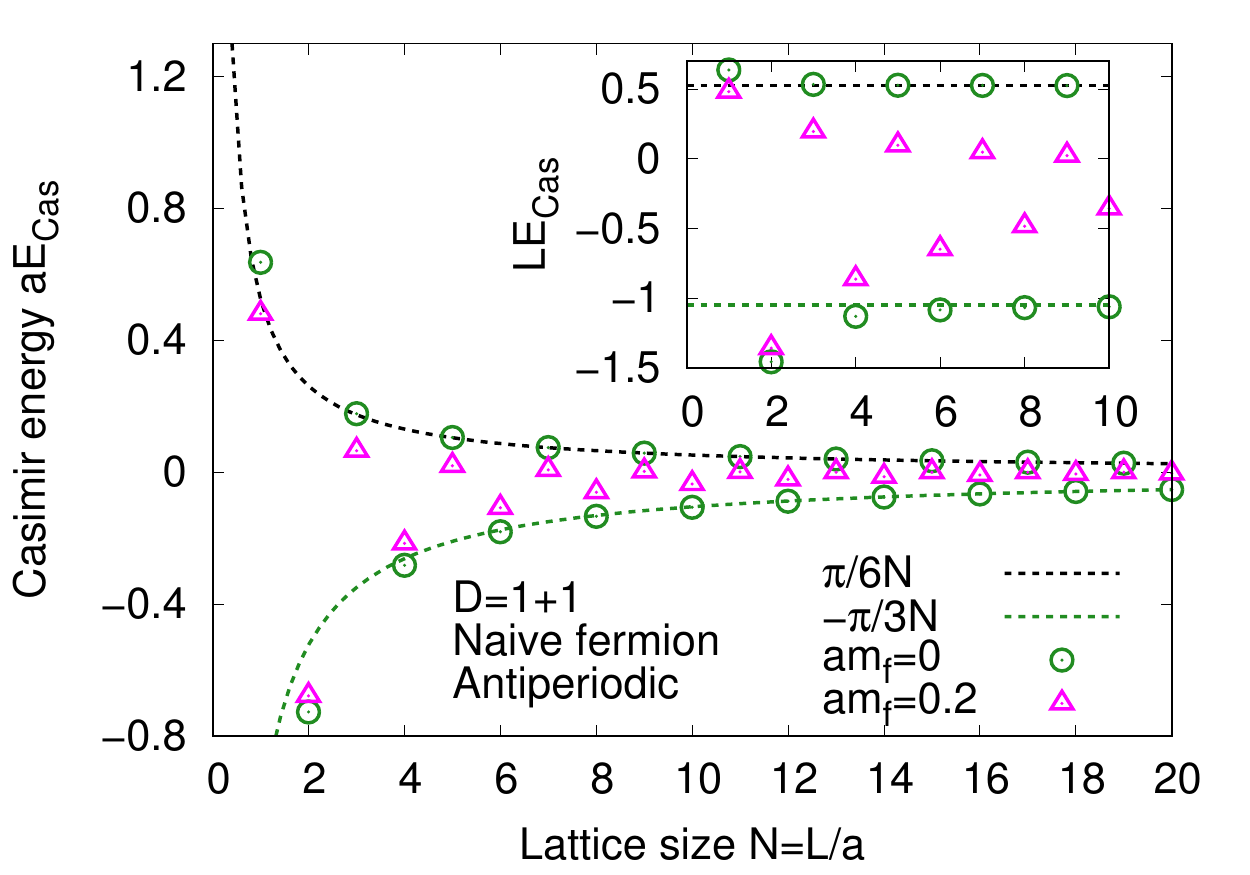}
        \end{center}
    \end{minipage}
    \caption{Casimir energy for massless or massive naive lattice fermion in the $1+1$ dimensional spacetime.
Upper: Periodic boundary.
Lower: Antiperiodic boundary.}
\label{fig:2d_nf}
\vspace{-10pt}
\end{figure}

In Fig.~\ref{fig:2d_nf}, we plot the Casimir energy for the massless and massive naive lattice fermions in the $1+1$ dimensional spacetime, where we plotted two dimensionless quantities: $a E_\mathrm{Cas}$ and $L E_\mathrm{Cas}$.
%Here, for the massless fermion, we compared the exact formulas~(\ref{eq:nf_exact1})--(\ref{eq:nf_exact4}) and the approximate formulas up to leading order~(\ref{eq:nf_aexp1})--(\ref{eq:nf_aexp4}).
From this figure, our findings are as follows:
\begin{enumerate}
\item {\it Oscillation of Casimir energy}---For both the periodic and antiperiodic boundaries, we find the oscillatory behavior of the Casimir energy.
This behavior is caused by the difference between the properties of the odd lattice and even lattice.
While, for the periodic boundary, the Casimir energy oscillates between two different repulsive forces, for the antiperiodic boundary, the Casimir energy oscillates between attractive and repulsive forces.

These behaviors can be qualitatively interpreted by the counting of the momentum zero modes in energy levels discretized by a finite volume.
For example, from the definition of the Brillouin zone, the discretized levels on the even lattice with the periodic boundary has two zero modes ($ap_1=0,\pi$).
Because of the appearance of zero modes, the vacuum energy in finite volume increases, which corresponds to the positive Casimir energy (and repulsive Casimir force).
Furthermore, the odd lattice with the periodic or antiperiodic boundary has one zero mode ($ap_1=0$ or $\pi$), which also leads to the repulsive Casimir energy. 
On the other hand, the even lattice with the antiperiodic boundary has no zero mode, so that the vacuum energy relatively decreases.
%This understanding is based on the simple energy level structures of the naive fermion, and for more complicated dispersion relations, a detailed discussion should be required.

\item {\it Equivalence between periodic and antiperiodic boundaries on odd lattice}---As seen in Eqs.~(\ref{eq:nf_exact1}) and (\ref{eq:nf_exact3}), we find that, on the odd lattice, the Casimir energy for the periodic and antiperiodic boundaries is equivalent.
This is because these dispersion relations are effectively equivalent to each other.

\item {\it Comparison with continuous Dirac fermions}---In the continuum theory with the one spatial dimension, the Casimir energy for free massless Dirac fermions is proportional to the inverse of the interval: $E_\mathrm{Cas} \propto 1/L$.
The Casimir energies for the periodic and antiperiodic boundary conditions are given by \cite{DePaola:1999im,Elizalde:2002wg}
\begin{align}
E_\mathrm{Cas}^\mathrm{2d,cont,P} = \frac{\pi}{3L}, \ \ \ \ E_\mathrm{Cas}^\mathrm{2d,cont,AP} = -\frac{\pi}{6L}, \label{eq:2dcont}
\end{align}
respectively.
Therefore, the Casimir energy for naive lattice fermions is completely different from that for the continuous Dirac fermion.
This disagreement remains even after the $a \to 0$ limit is taken.
\item {\it Dependence on lattice spacing}---In Fig.~\ref{fig:2d_nf}, we compare the approximated formulas with the $a/L$ expansion, Eqs.~(\ref{eq:nf_aexp1})--(\ref{eq:nf_aexp4}), and the exact formulas (\ref{eq:nf_exact1})--(\ref{eq:nf_exact4}).
These formulas are in good agreement with each other in the region of $N \geq 3$, which indicates that the approximated formulas will be useful for estimating the Casimir energy as far as we focus on $N \geq 3$.
\item {\it Suppression of massive Casimir energy}---In Fig.~\ref{fig:2d_nf}, we compare the results of the massless fermion and massive one with $a m_f=0.2$.
When the fermion has a nonzero mass, the Casimir energy is suppressed.
This tendency is similar to the usual Casimir energy with massive degrees of freedom.
We emphasize that although the Casimir energy is suppressed, the oscillatory behavior survives. 
\end{enumerate}

%%%%%%%%%%%%%%%%%%%%%%%%%%%%%%%%
\section{Casimir energy for Wilson fermion} \label{Sec_4}
%%%%%%%%%%%%%%%%%%%%%%%%%%%%%%%%
Next, we investigate the properties of Casimir effects for Wilson fermions.
We define the Wilson Dirac operator $D_\mathrm{W}$ with the Wilson coefficient $r$:
\begin{align}
aD_\mathrm{W} \equiv i \sum_k \gamma_k \sin ap_k + r \sum_k (1 - \cos ap_k) + am_f. \label{eq:Dw}
\end{align}
The dispersion relation is
\begin{align}
a^2 E_{\mathrm{W}}^2(ap) = \sum_k \sin^2 ap_k + \left[ r \sum_k (1 - \cos ap_k) + am_f \right]^2. \label{eq:disp_w}
\end{align}
From Eq.~(\ref{eq:def_cas}), we can calculate the Casimir energy.
For the Wilson fermion with $r=1$ and $am_f=0$ in the $1+1$ dimensional spacetime, we can get the exact formulas:
\begin{align}
aE_\mathrm{Cas}^\mathrm{2d,W,P} &= \frac{4N}{\pi} - 2\cot \frac{\pi}{2N}, \label{eq:Wil_exactP} \\
aE_\mathrm{Cas}^\mathrm{2d,W,AP} &= \frac{4N}{\pi} - 2\csc \frac{\pi}{2N}. \label{eq:Wil_exactAP}
\end{align}
When we expand these formulas by $a/L$, we obtain
\begin{align}
E_\mathrm{Cas}^\mathrm{2d,W,P} &= \frac{\pi}{3L} +\frac{\pi^3 a^2}{180L^3} + \mathcal{O}(a^4),  \\
E_\mathrm{Cas}^\mathrm{2d,W,AP} &= -\frac{\pi}{6L} -\frac{7\pi^3 a^2}{1440L^3} + \mathcal{O}(a^4).
\end{align}

\begin{figure}[t!]
    \begin{minipage}[t]{1.0\columnwidth}
        \begin{center}
            \includegraphics[clip, width=1.0\columnwidth]{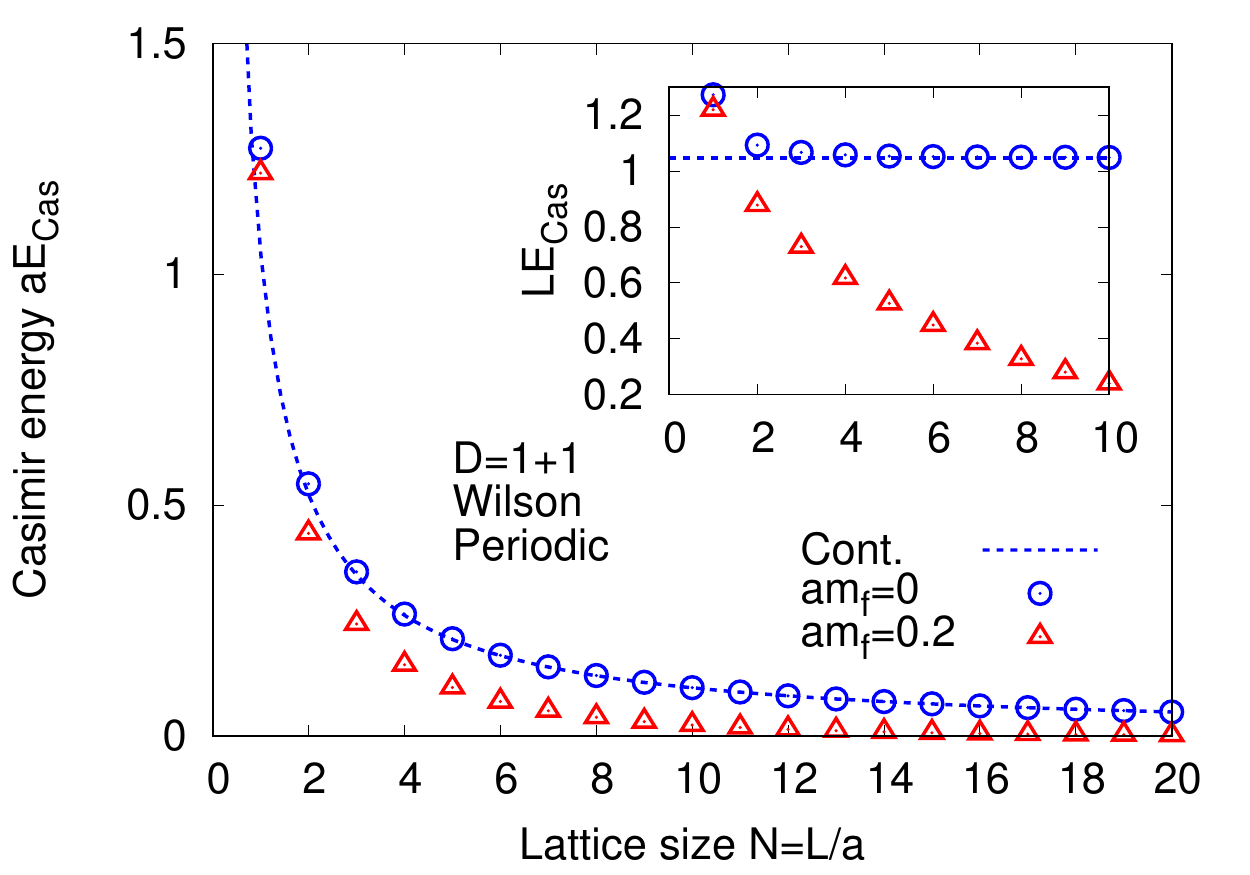}
        \end{center}
    \end{minipage}
    \begin{minipage}[t]{1.0\columnwidth}
        \begin{center}
            \includegraphics[clip, width=1.0\columnwidth]{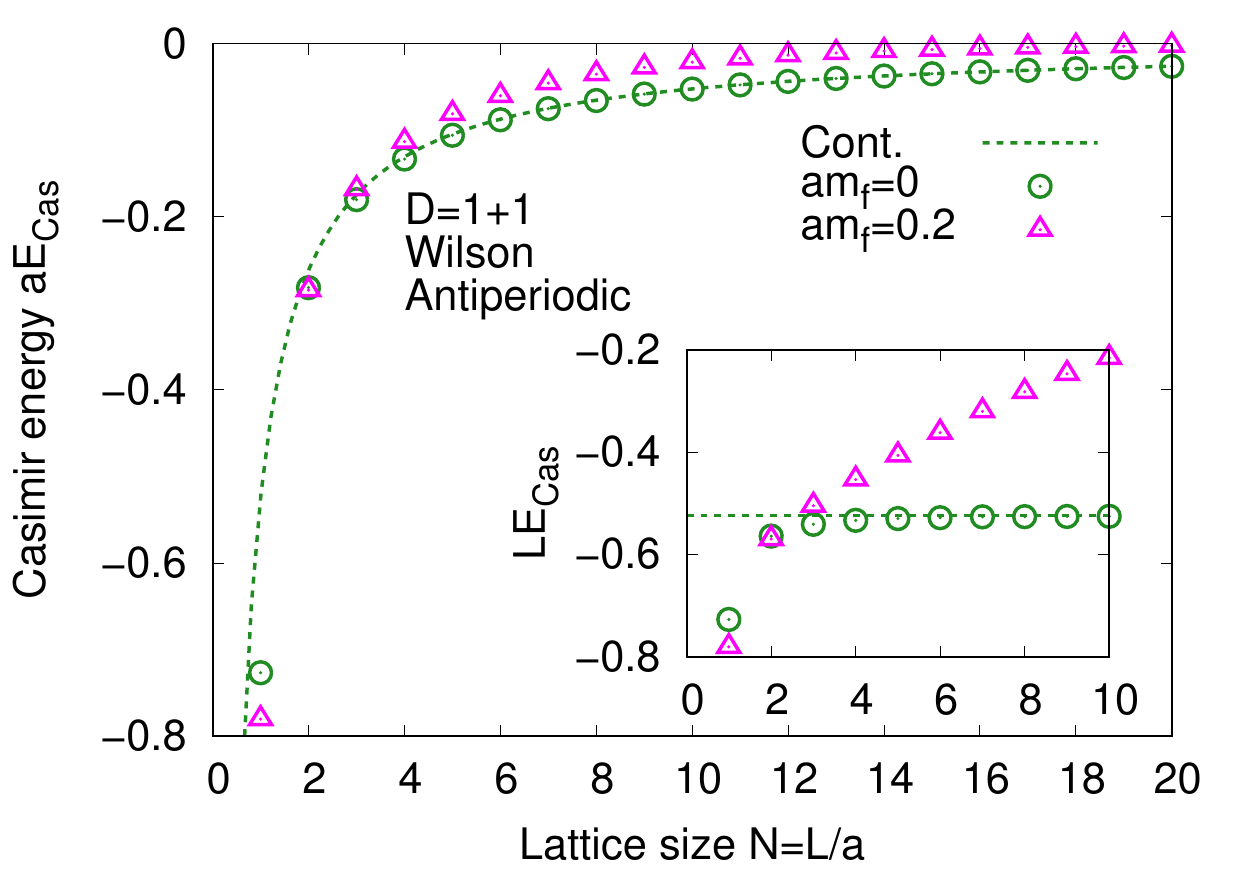}
        \end{center}
    \end{minipage}
    \caption{Casimir energy for massless or massive Wilson fermion in the $1+1$ dimensional spacetime.
Upper: Periodic boundary.
Lower: Antiperiodic boundary.}
\label{fig:2d_Wil}
\vspace{-10pt}
\end{figure}

In Fig.~\ref{fig:2d_Wil}, we show the results of the Wilson fermion at $r=1$ in the $1+1$ dimensional spacetime.
From this figure, our findings are as follows:
\begin{enumerate}
\item{\it Agreement with continuum theory in large size}---In the larger lattice size with $N \geq 3$, the Casimir energy for the Wilson fermion agrees very well the Casimir energies (\ref{eq:2dcont}) in the continuum theory.
This result indicates that when one tries to measure the fermionic Casimir energy in this region by using numerical simulations, one can obtain the results consistent with the continuum theory within a small discretization error.
In particular, the oscillation of the Casimir energy found for the naive fermion is removed by the Wilson term.
In other words, the continuum limit can be easily taken.

Moreover, we emphasize that our procedure with the $a \to 0$ limit is nothing but a new derivation of the fermionic Casimir energy known in the continuum theory.
In our approach, the divergence of the vacuum energy is successfully regularized by the lattice regularization with the Wilson fermion, without using conventional mathematical approaches such as the zeta function regularization and Abel-Plana formulas.

\item{\it Overestimate in small size}---Furthermore, we find that, in the smaller lattice size of $N \leq 3$, the Casimir energy for the Wilson fermion is overestimated by the discretization effects.
This is because the energy-momentum dispersion relation of the Wilson fermions is underestimated compared to that of the Dirac fermion in the continuum theory.

Thus, the Casimir energy for the Wilson fermion is larger than that of the Dirac fermion.
Therefore, in order to observe the Casimir energy for lattice fermions, the material with the Wilson-fermion-like band structure will be more preferable.
\item{\it Massive Casimir energy}---In the massive case, we find that the Casimir energy is suppressed for $N \geq 2$, which is consistent with the usual suppression of the Casimir effect by massive degrees of freedom.
On the other hand, for $N=1$ with the antiperiodic boundary, we find the enhancement of Casimir energy by including a nonzero mass.
This is because the mode allowed at $N=1$ is only $ap_1=\pi$, and then the discretized energy level is dominated by $aE = 2+am_f$.
%As a result, such a single ``heaviest" mode induces a Casimir energy more attractive than the massless case.
%This situation is different from the periodic boundary, where the discretized energy level is dominated by $ap_1=0$ and $aE = am_f$.
%Thus, we emphasize that {\it the Casimir effect enhanced by a nonzero mass} is a rare example in the long history of the Casimir effect.
\end{enumerate}

%%%%%%%%%%%%%%%%%%%%%%%%%%%%%%%%
\section{Casimir energy for overlap fermion} \label{Sec_5}
%%%%%%%%%%%%%%%%%%%%%%%%%%%%%%%%
In this section, we investigate the Casimir energy of the overlap fermion with the MDW kernel operator.
In the domain-wall fermion formulation~\cite{Kaplan:1992bt,Shamir:1993zy,Furman:1994ky}, a ``bulk" fermion defined as kernel operators in the $D+1$-dimensional Euclidean space is projected into the chiral ``surface" fermion in the $D$-dimensional Euclidean space.
The length of the extra dimension is usually finite, but we assume infinite length for simplicity, which corresponds to the overlap fermion~\cite{Neuberger:1997fp,Neuberger:1998wv}.

%\subsection{MDW kernel operators}
We define the MDW kernel operator $D_\mathrm{MDW}$ using the M\"obius parameters $b$ and $c$~\cite{Brower:2004xi,Brower:2005qw,Brower:2012vk},
\begin{align}
aD_\mathrm{MDW} \equiv \frac{b (aD_\mathrm{W})}{2 + c (aD_\mathrm{W})}.
\end{align}
This kernel operator is a generalization of the conventional Shamir-type ($b=1,c=1$)~\cite{Shamir:1993zy} and Bori\c{c}i-type (or Wilson-type) ($b=2,c=0$)~\cite{Borici:1999zw,Borici:1999da} formulations.
$D_\mathrm{W}$ is the Wilson Dirac operator defined as Eq.~(\ref{eq:Dw}) with $r=1$, but the fermion mass $m_f$ is replaced by the domain-wall height $am_f \to -M_0$ which is a negative mass.

%\subsection{Overlap fermion with MDW kernel}
Using the MDW kernel operator $D_\mathrm{MDW}$, we introduce the overlap Dirac operator $D_\mathrm{OV}$ with fermion mass $m_f$,
\begin{align}
aD_\mathrm{OV}
\equiv
(2 - cM_0)
M_0
m_\mathrm{PV} \frac{(1 + am_f) + (1 - am_f)V }
{(1 + m_\mathrm{PV}) + (1 - m_\mathrm{PV})V},
\end{align}
with $V \equiv \gamma_5\mathrm{sign}(\gamma_5 a D_\mathrm{MDW}) = D_\mathrm{MDW}/\sqrt{D_\mathrm{MDW} ^\dagger D_\mathrm{MDW}}$.
%\begin{align}
%V \equiv \gamma_5\mathrm{sign}(\gamma_5 a D_\mathrm{MDW}) = \frac{D_\mathrm{MDW}}{\sqrt{D_\mathrm{MDW} ^\dagger D_\mathrm{MDW}}}.
%\end{align}
The Pauli-Villars mass $m_\mathrm{PV}$ was introduced so as to satisfy the Ginsparg-Wilson relation.
The scaling factor $
(2 - cM_0) M_0 m_\mathrm{PV}$ with a constraint $2 - cM_0 > 0$ is determined so as to realize the dispersion relation of fermions in the continuum theory: $\lim_{a \rightarrow 0} D^\dagger _\mathrm{OV}D_\mathrm{OV} = p^2$ for $m_f = 0$.

The dispersion relation for the overlap fermion is
\begin{align}
a^2E^2_\mathrm{OV}
& =
\left[(2 - cM_0)
M_0 m_\mathrm{PV} \right]^2 \nonumber\\
& \times \frac{2[1 + (am_f)^2] + [1 - (am_f)^2] (V^\dagger + V)}
{2[1 + m_\mathrm{PV}^2] + [1 - m_\mathrm{PV}^2](V^\dagger + V)}. \label{eq:dispOV}
\end{align}
%where we used the $V^\dagger V = 1$ and the commutation relation between $V^\dagger + V $ and $V$.
%$V^\dagger + V$ are represented by the Wilson operator $D_\mathrm{W}$,
%\begin{align}
%V^\dagger + V
%&=
%\left(D_\mathrm{MDW} ^\dagger + D_\mathrm{MDW} \right)\frac{1}{\sqrt{D_\mathrm{MDW} ^\dagger D_\mathrm{MDW}}}, \notag\\
%&=
%\frac{1}
%{\sqrt{D_\mathrm{W} ^\dagger D_\mathrm{W}}}
%\frac{2(D_\mathrm{W} + D_\mathrm{W} ^\dagger + cD_\mathrm{W}^\dagger D_\mathrm{W})}{\sqrt{{4 + 2c(D_\mathrm{W} ^\dagger + D_\mathrm{W}) + c^2 D_\mathrm{W} ^\dagger D_\mathrm{W}}}}, \label{eq:Vdag+V}
%\end{align}
%where we used the properties $D_\mathrm{W}^\dagger D_\mathrm{W} > 0$ and ${4 + 2c(D_\mathrm{W} ^\dagger + D_\mathrm{W}) + c^2 D_\mathrm{W} ^\dagger D_\mathrm{W}} > 0$.
Note that, in this form, the parameter $b$ dependence is completely eliminated.
In this work, the $b$ dependence is not relevant since we consider the infinite length of the extra dimension.
From the dispersion relation (\ref{eq:dispOV}) and the definition (\ref{eq:def_cas}), we can calculate the Casimir energy of the overlap fermion.

%\subsection{Numerical results}
\begin{figure}[t!]
    \begin{minipage}[t]{1.0\columnwidth}
        \begin{center}
            \includegraphics[clip, width=1.0\columnwidth]{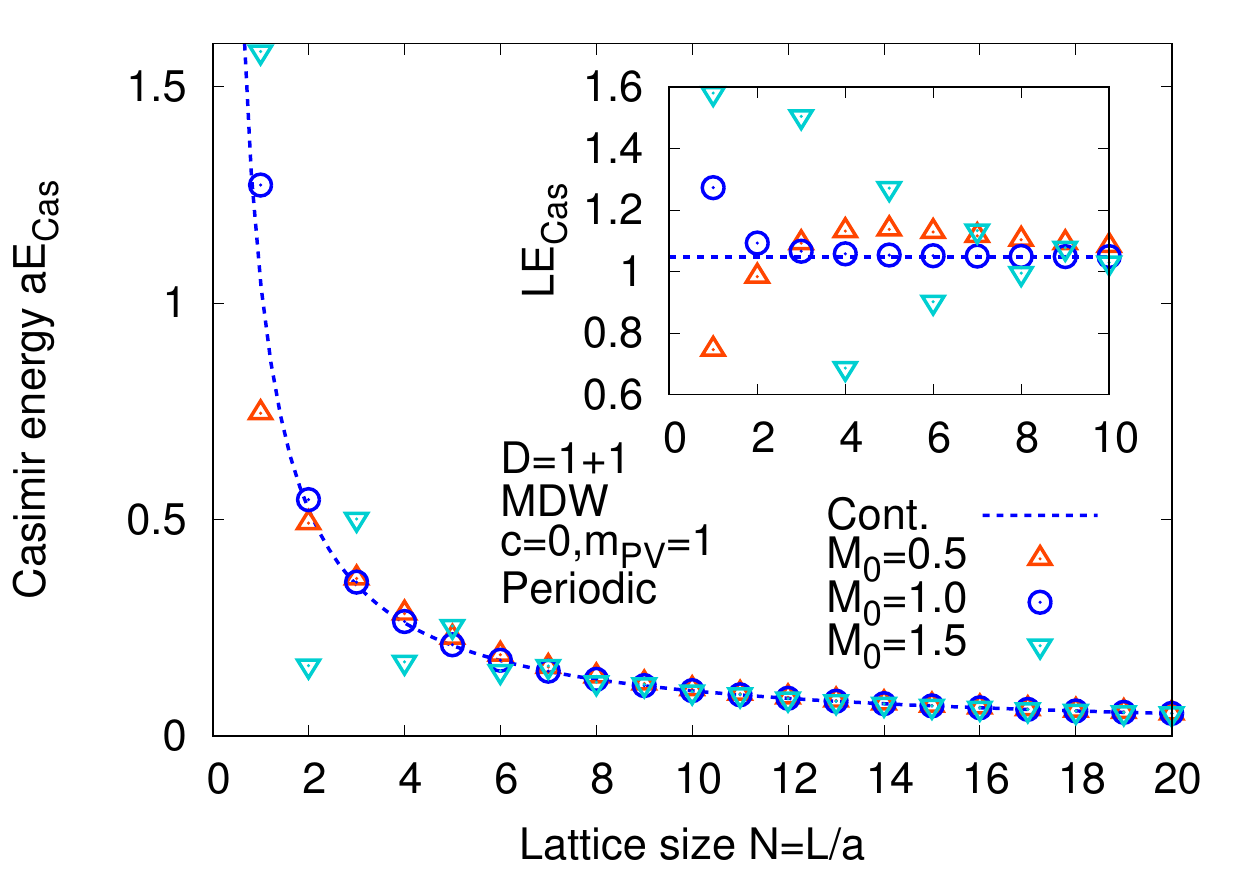}
            \includegraphics[clip, width=1.0\columnwidth]{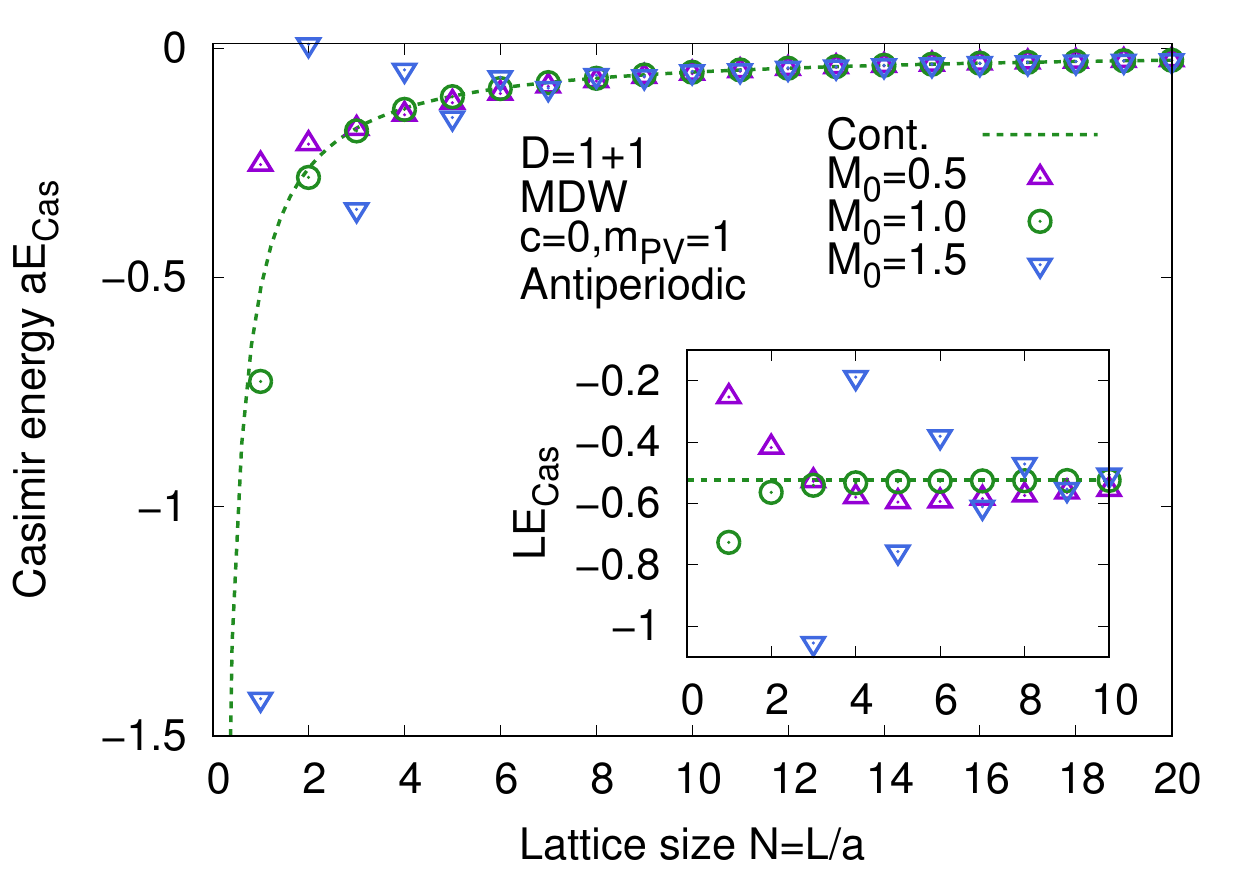}
        \end{center}
    \end{minipage}
    \caption{Domain-wall height $M_0$ dependence of Casimir energy for overlap fermions with MDW kernel operator in the $1+1$ dimensional spacetime.
Upper: Periodic boundary. Lower: Antiperiodic boundary.}
\label{fig:2d_M0}
\vspace{-10pt}
\end{figure}

\begin{figure}[t!]
    \begin{minipage}[t]{1.0\columnwidth}
        \begin{center}
            \includegraphics[clip, width=1.0\columnwidth]{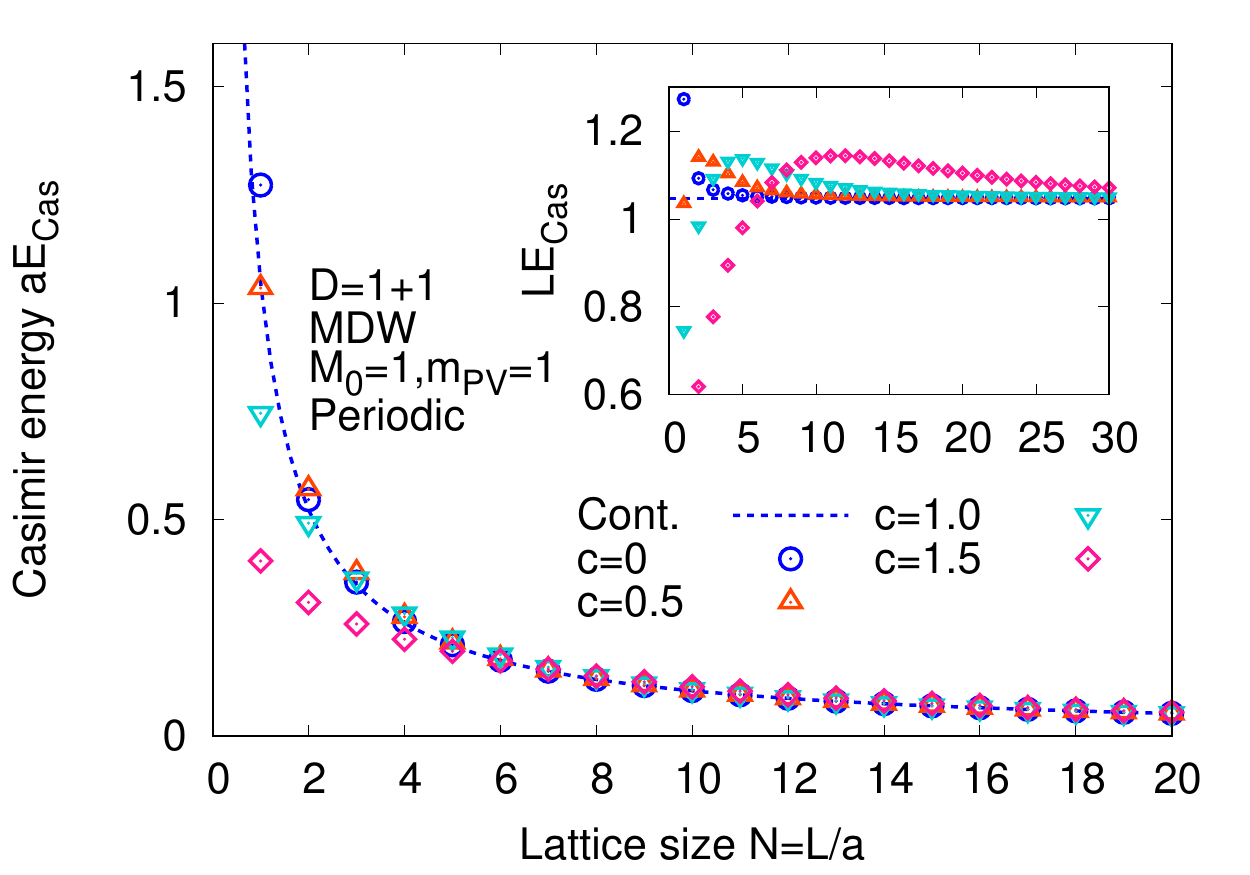}
            \includegraphics[clip, width=1.0\columnwidth]{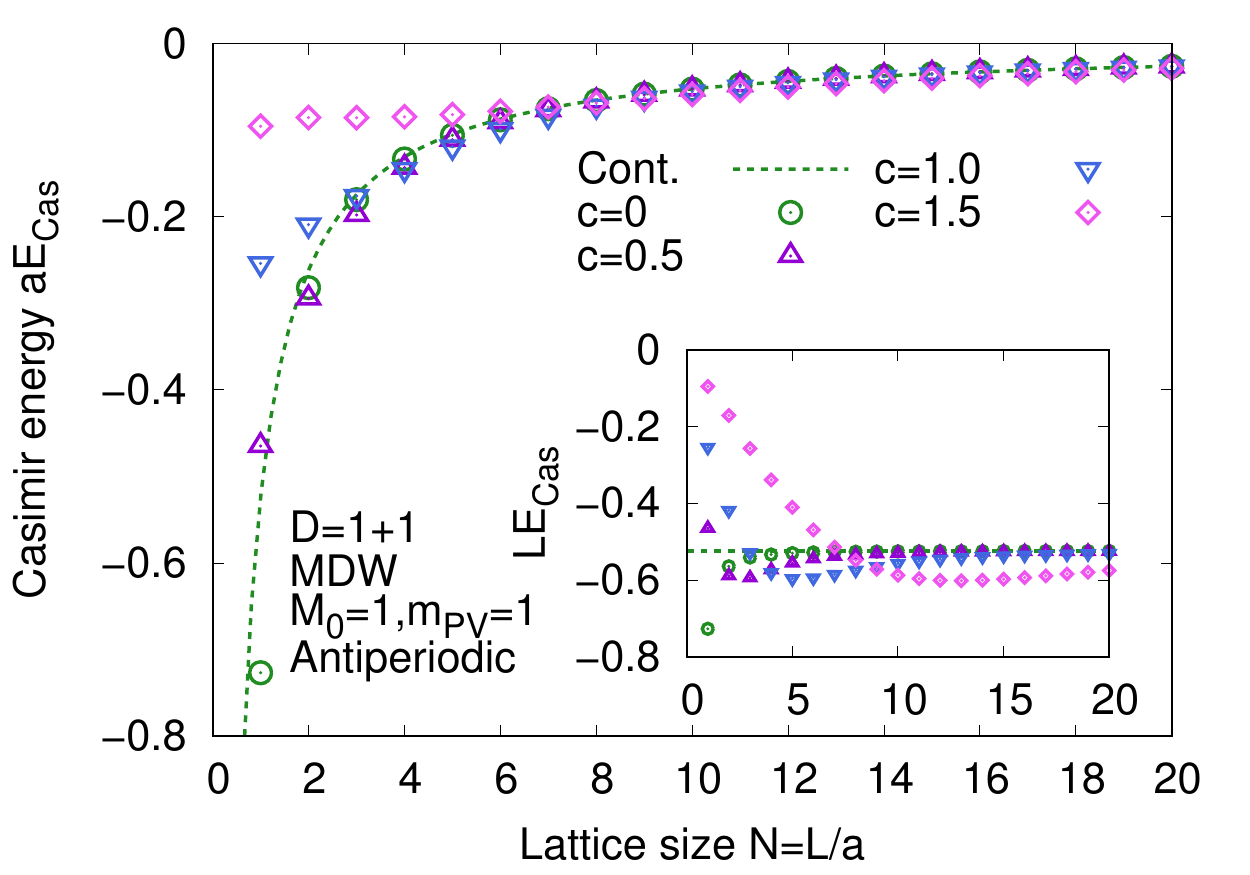}
        \end{center}
    \end{minipage}
    \caption{M\"obius parameter $c$ dependence of Casimir energy for overlap fermions with MDW kernel operator in the $1+1$ dimensional spacetime.
Upper: Periodic boundary. Lower: Antiperiodic boundary.}
\label{fig:2d_c}
\vspace{-10pt}
\end{figure}

\begin{figure}[tb!]
    \begin{minipage}[t]{1.0\columnwidth}
        \begin{center}
            \includegraphics[clip, width=1.0\columnwidth]{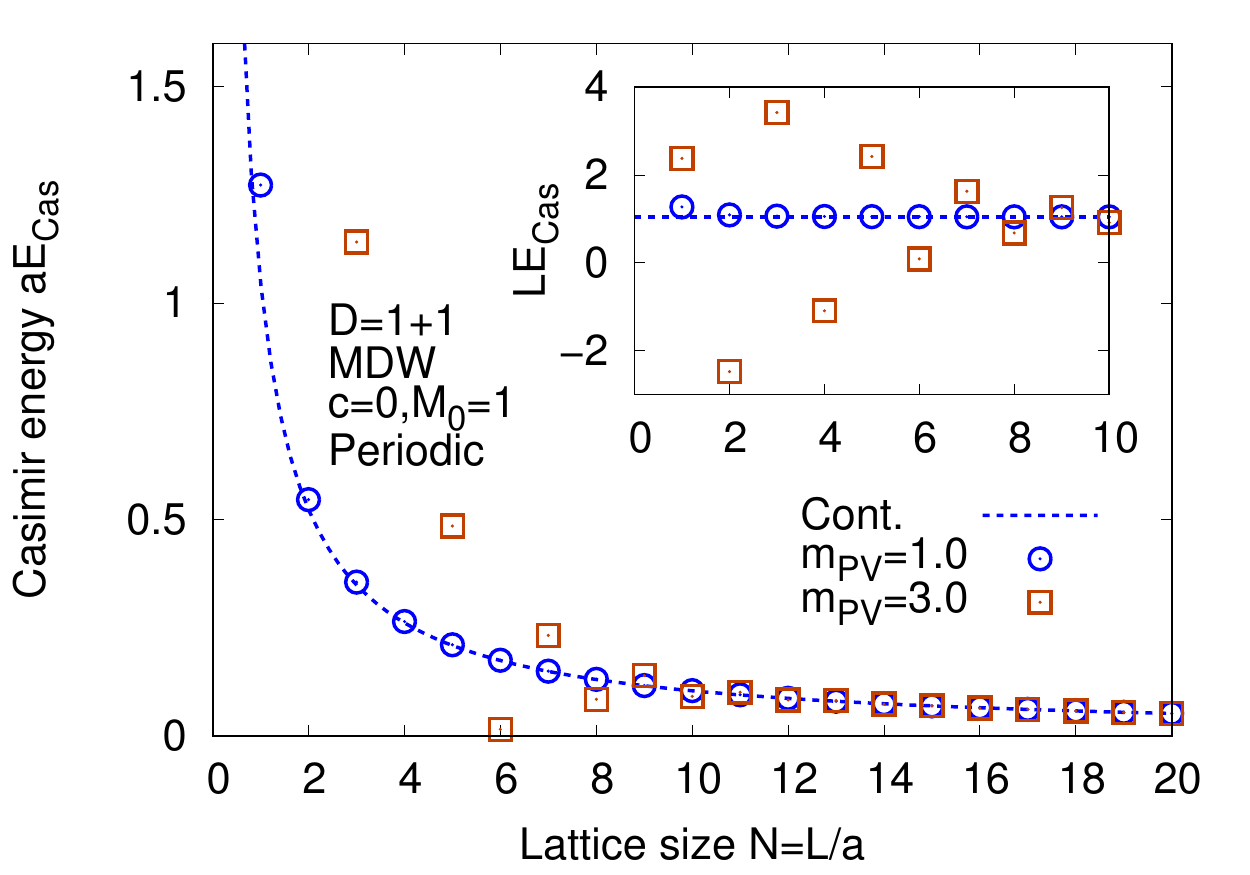}
            \includegraphics[clip, width=1.0\columnwidth]{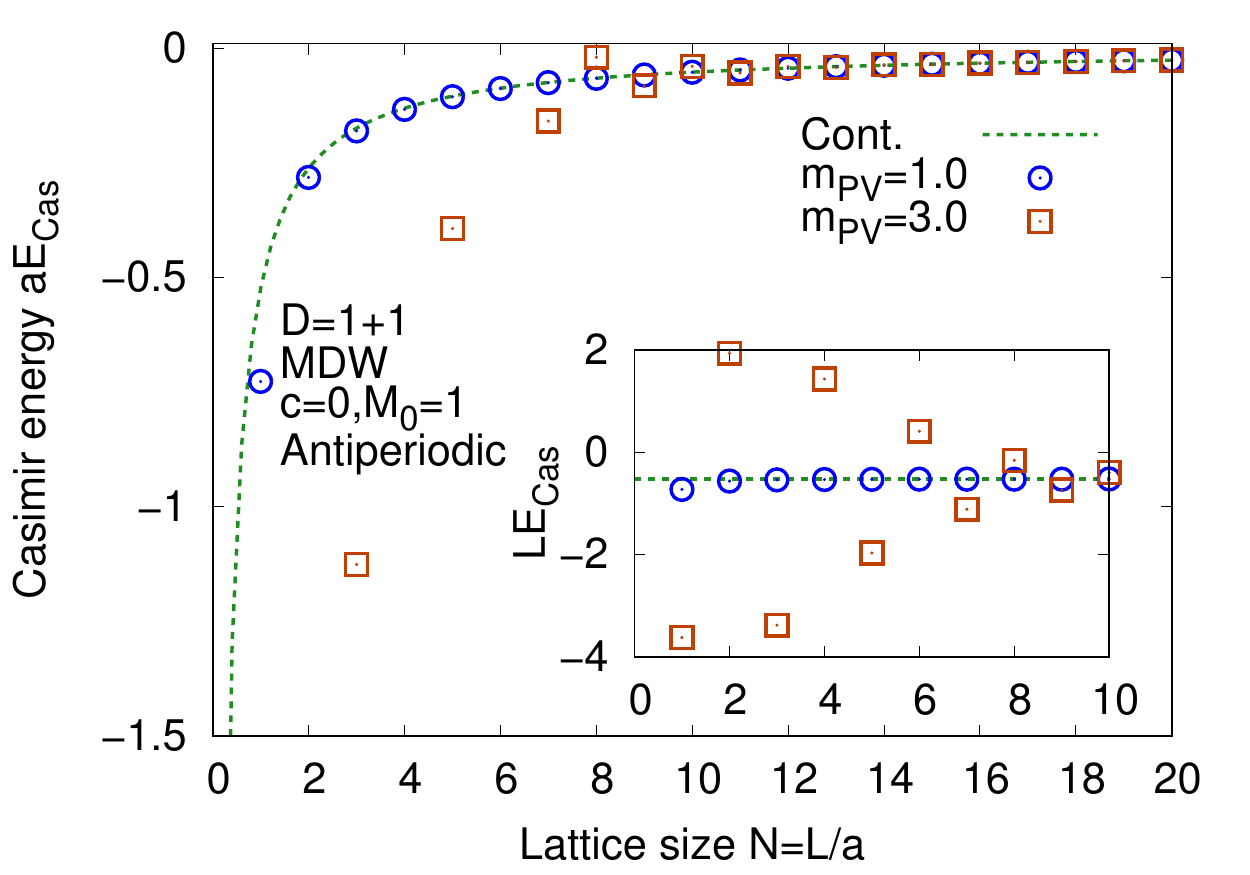}
        \end{center}
    \end{minipage}
    \caption{Pauli Villars mass $m_\mathrm{PV} $ dependence of Casimir energy for overlap fermions with MDW kernel operator in the $1+1$ dimensional spacetime.
Upper: Periodic boundary. Lower: Antiperiodic boundary.}
\label{fig:2d_mpv}
\vspace{-10pt}
\end{figure}

In Fig.~\ref{fig:2d_M0}, we show the dependence of the Casimir energy for the overlap fermion on the domain-wall height ($M_0 =0.5$, $1.0$, and $1.5$) at fixed $c=0$ and $m_\mathrm{PV}=1.0$.
Among them, the Casimir energy at $M_0=1.0$ well reproduces that in the continuum theory, which is equivalent to that for the Wilson fermion.
This equivalence is easily derived by Eqs.~(\ref{eq:disp_w}) and (\ref{eq:dispOV}).
At $M_0=0.5$, we find the overestimate of the Casimir energy in the larger lattice size and underestimate in the smaller lattice size.
At $M_0=1.5$, there is an oscillatory behavior for both the periodic and antiperiodic boundaries, where we find the overestimate on the odd lattice and underestimate on the even lattice.
This oscillation is induced by the appearance of massive doublers, which is absent at $M_0 \leq 1.0$.
The setup with $M_0>1.0$ is practically important in lattice simulations with gauge fields (e.g., see Refs.~\cite{Blum:2000kn,Aoki:2002vt,Brower:2004xi,Brower:2005qw,Brower:2012vk}) because such $M_0$ may useful for correcting the fermion dispersion relations effectively modified by gauge fields.
A similar oscillation is also found at finite temperature~\cite{Banerjee:2008ii,Gavai:2008ea}.

In Fig.~\ref{fig:2d_c}, we show the dependence of the Casimir energy on M\"obius parameter ($c=0$, $0.5$, $1.0$, and $1.5$) at fixed $M_0=1.0$ and $m_\mathrm{PV}=1.0$.
We find that the Casimir energy at $c=0$, which is equivalent to the Wilson fermion, best reproduces that in the continuum theory.
For a nonzero $c$, we find the overestimate of the Casimir energy in the larger lattice size and underestimate in the smaller lattice size.
As $c$ is larger, the region with deviation from the continuum theory becomes broader.
In particular, at $c=1.5$, this deviation can be observed even on the lattice with $N = 30$.

In Fig.~\ref{fig:2d_mpv}, we show the dependence of the Casimir energy on the Pauli Villars mass $m_\mathrm{PV} $ dependence ($m_\mathrm{PV} =1.0$ and $3.0$).
At $m_\mathrm{PV}=3.0$, we find an oscillatory behavior.
This oscillation is also induced by the appearance of massive doublers, which is similar to $M_0 > 1.0$.
A heavy Pauli Villars mass, $m_\mathrm{PV} > 1.0$, could be practically useful in lattice simulations~\cite{Nakayama:2018ubk}.

%%%%%%%%%%%%%%%%%%%%%%%%%%%%%%%%
\section{Conclusion and outlook} \label{Sec_6}
%%%%%%%%%%%%%%%%%%%%%%%%%%%%%%%%
In this Letter, we defined the Casimir energy for lattice fermions, Eq.~(\ref{eq:def_cas_4d}), for the first time.
From this definition, we investigated the properties of the Casimir energy for the massless/massive naive fermion, Wilson fermion, and overlap fermion with the MDW kernel operator.
For some types of fermions, we found a characteristic oscillatory behavior between odd and even lattices.
%\footnote{
%The oscillatory phenomena of Casimir energy were also found in other contexts~\cite{Fuchs:2007,Wachter:2007,Kolomeisky:2008,Zhabinskaya:2009}.
%}

For some simple cases, we analytically obtained the exact formulas for the Casimir energies, and for more complicated cases, we numerically calculated the Casimir energies.
In order to carefully examine our formulas, the confirmation from other mathematical derivation is left for future works.

In this Letter, we focused on only free fermions, but the Casimir energy for interacting lattice fermions would be also important.
In particular, one might be interested in the relation between the Casimir effect and the parity-broken (Aoki) phase \cite{Aoki:1983qi} realized in the strong-coupling regime of the Wilson fermion and domain-wall fermion.
The original Casimir effect in the QED vacuum is dominated by the photon field, and the contribution from fermions (mainly, electrons) is regarded as higher-loop corrections of the photon self-energy in the perturbation theory of QED.
Our findings are related to the comprehensive understandings of the Casimir effect for the lattice QED or QCD.
To study the contribution from an interaction between a lattice fermion and a gauge field, analyses with dispersion relations modified by the interaction would be interesting.

Other fermion actions not studied in this Letter, such as staggered fermions~\cite{Kogut:1974ag,Susskind:1976jm}, will be interesting.
Also, we showed only the results in the $1+1$ dimensional spacetime, but the application to the $2+1$ and $3+1$ dimensions is straightforward, where the oscillatory behavior also appears~\cite{Ishikawa:2020}.

Casimir effects for lattice fermions will be observed in tabletop experiments with Dirac electron systems {\it when an extremely small lattice size is realized}.
For example, a (very short) 1D ring made of a Dirac material can induce the Casimir energy with the periodic boundary condition in the one-dimensional space.
Similarly, a (very small) cylinder made of 2D thin films leads to the Casimir energy in the two-dimensional space (for works related to carbon nanotubes, see Refs.~\cite{Bellucci:2009jr,Bellucci:2009hh}).
In this sense, to investigate fermionic Casimir effects in honeycomb lattices will be important.
Another possible candidate for studying the Casimir effects for lattice fermions would be cold-atom simulations \cite{Bermudez:2010da,Mazza:2011kf,Kuno:2018rmc,Zache:2018jbt} with a small size.

%%%%%%%%%%%%%%%%%%%%%%%%%%%%%%%%
\section*{Acknowledgment}
%%%%%%%%%%%%%%%%%%%%%%%%%%%%%%%%
The authors are grateful to Yasufumi Araki and Daiki Suenaga for giving us helpful comments.
This work was supported by Japan Society for the Promotion of Science (JSPS) KAKENHI (Grant Nos. JP17K14277 and JP20K14476).

%%%%%%%%%%%%%%%%%%%%%%%%%%%%%%%%
\appendix
%%%%%%%%%%%%%%%%%%%%%%%%%%%%%%%%
\bibliography{casimir_ref}
%%%%%%%%%%%%%%%%%%%%%%%%%%%%%%%%
\end{document}